# CALA-*n*: A Quantum Library for Realizing Cost-Effective 2-, 3-, 4-, and 5-bit Gates on IBM Quantum Computers using Bloch Sphere Approach, Clifford+T Gates, and Layouts

Ali Al-Bayaty
*Electrical and Computer Eng. Dept.*
*Portland State University*
*Oregon, USA*
albayaty@pdx.edu

Xiaoyu Song
*Electrical and Computer Eng. Dept.*
*Portland State University*
*Oregon, USA*
songx@pdx.edu

Marek Perkowski
*Electrical and Computer Eng. Dept.*
*Portland State University*
*Oregon, USA*
h8mp@pdx.edu

*Abstract*—We introduce a new quantum layout-aware approach to realize cost-effective *n*-bit gates using the Bloch sphere, for $2 \leq n \leq 5$ qubits. These *n*-bit gates are entirely constructed from the Clifford+T gates, in the approach of selecting sequences of rotations visualized on the Bloch sphere. This Bloch sphere approach ensures to match the quantum layout for synthesizing (transpiling) these *n*-bit gates into an IBM quantum computer. Various standard *n*-bit gates (Toffoli, Fredkin, etc.) and their operational equivalent of our proposed *n*-bit gates are examined and evaluated, in the context of the final quantum costs, as the final counts of generated IBM native gates. In this paper, we demonstrate that all our *n*-bit gates always have lower quantum costs than those of standard *n*-bit gates after transpilation. Hence, our Bloch sphere approach can be used to build a quantum library of various cost-effective *n*-bit gates for different layouts of IBM quantum computers.

**Keywords:** *Bloch sphere, CALA-n, Clifford+T gates, IBM quantum computers, quantum layouts, transpilation*

## 1 INTRODUCTION

A quantum gate, in general, is classified into a single-qubit, double-qubit, or multiple-qubit gate. A single-qubit gate directly performs its quantum operation on a single qubit. For a double-qubit gate, one of its two qubits operates the quantum operation, while the other qubit represents the output of this quantum operation. Whileas for a multiple-qubit gate, $n-1$ qubits operate the quantum operation, and the remaining qubit represents the output of this quantum operation, where $n \geq 3$. For that, the operating qubits are termed the "control qubits", and the output qubit is termed the "target qubit". To perform the quantum operation of a multiple-qubit (or a double-qubit) gate on its target qubit, all control qubits should be set to the |1⟩ state, except for the following two quantum gates:

- The SWAP (as a double-qubit) gate has two target qubits with no control qubits specified. This gate always switches the indices of its two input qubits as swapped target qubits.
- The controlled-SWAP (as a multiple-qubit) gate has one control qubit and two target qubits. This gate performs the same switching operation as the SWAP gate, when its control qubit sets to the |1⟩ state. The controlled-SWAP gate is the so-called "Fredkin gate".

In addition, the quantum operations of single-qubit gates are categorized based on their rotations around the three axes (X-axis, Y-axis, and Z-axis) of the Bloch sphere [1-3]. Hence, the names of single-qubit gates are derived from their rotational operations around these three axes, e.g., the Pauli-X (X), Pauli-Y (Y), and Pauli-Z (Z) single-qubit gates rotate around the X-axis, Y-axis, and Z-axis of the Bloch sphere, respectively, in $\pi$ radians. However, fractions of $\pi$ are allowed to rotate single-qubit gates around their corresponding axes of the Bloch sphere, and accordingly such gates are termed the rotational-X (as RX($\alpha$)), rotational-Y (as RY($\beta$)), and rotational-Z (as RZ($\gamma$)) gates, where $\alpha$, $\beta$, and $\gamma$ are the rotational angles range between [0, $\pm 2\pi$]. Note that the RX(+ $\pi/2$), RX(– $\pi/2$), RZ(+ $\pi/2$), RZ(– $\pi/2$), RZ(+ $\pi/4$), and RZ(– $\pi/4$) single-qubit gates are also called the $\sqrt{X}$ (or V), $\sqrt{X}^\dagger$ (or V$^\dagger$), $\sqrt{Z}$ (or S), $\sqrt{Z}^\dagger$ (or S$^\dagger$), $\sqrt[4]{Z}$ (or T), and $\sqrt[4]{Z}^\dagger$ (or T$^\dagger$) gates, respectively [1-4]. Note that (i) the + and – signs indicate the counterclockwise and clockwise rotations around an axis of the Bloch sphere, respectively, (ii) the identity (I) gate is a single-qubit gate that does not perform any quantum operation, i.e., I does not rotate around any axis of the Bloch sphere, and (iii) the Hadamard (H) gate is a single-qubit gate that performs three quantum operations in the sequence of three single-qubit gates {S $\sqrt{X}$ S}, i.e., H half-rotates once around the X-axis and half-rotates twice around the Z-axis of the Bloch sphere.

When a number of control qubits are applied to a single-qubit X gate, the double-qubit or multiple-qubit gate is then constructed, e.g., the controlled-X (CX), controlled-controlled-X (CCX), or ($n$–1)-controlled-X (C$^{n-1}$X) gate, where $n \geq 3$ qubits. In quantum computing, the CX, CCX, and C$^{n-1}$X gates are the so-called "Feynman", "Toffoli (or 3-bit Toffoli)", and "$n$-bit Toffoli" gates, respectively. Note that, in this paper, we used the term "bit" to represent a qubit, as proposed by Barenco *et al.* [4]. Similarly, the double-qubit and multiple-qubit gates from single-qubit Y and Z gates can be constructed as well, e.g., the controlled-Y (CY), controlled-Z (CZ), ($n$–1)-controlled-Y (C$^{n-1}$Y), and ($n$–1)-controlled-Z (C$^{n-1}$Z) gates, where the C$^{n-1}$Y and C$^{n-1}$Z gates are also called the "multiple-controlled-Y (MCY)" and "multiple-controlled-Z (MCZ)" gates, respectively.

The quantum operations of all above-mentioned single-qubit, double-qubit, and multiple-qubit gates can be either (i) expressed mathematically using the unitary matrices [1-4], as the numerical representation of their quantum rotational operations, or (ii) represented geometrically using the Bloch sphere [1-3], as the visual representation of transitions for the states of a qubit after a gate (or a sequence of gates) applies on it.

A specific group of single-qubit and double-qubit gates form the so-called "Clifford+T gates" [5-7], which consists of: I, X, Y, Z, H, $\sqrt{X}$, $\sqrt{X}^\dagger$, S, S$^\dagger$, T, T$^\dagger$, CX, CY, CZ, and SWAP. The Clifford+T gates are an important aspect of quantum computing, since they (i) invert the phase of a Pauli gate, e.g., {Z X Z} = – X, and (ii) construct Pauli gates that a quantum processing unit (QPU) does not support them, e.g., a QPU only supports the X, S, and S$^\dagger$ gates, but it does not support the Y gate, then {S X S$^\dagger$} = Y. A QPU can support a limited set of single-qubit and double-qubit gates, which are the so-called "native gates" or "basis gates". For instance, the `ibm_brisbane` QPU [8, 9] of 127 qubits only supports the native gates of I, X, $\sqrt{X}$, RZ, and ECR (echoed cross-resonance). Since this QPU does not support all Clifford+T gates, then certain transformations for these native gates are required to construct other Clifford+T gates, e.g., the H gate is constructed using the transformation of {RZ($\pi/2$) $\sqrt{X}$ RZ($\pi/2$)} = {S $\sqrt{X}$ S}.

Note that the ECR gate is the only double-qubit native gate for most IBM QPUs of 127 qubits, since the IBM quantum system replaces its native CX gate with the ECR gate to: (i) minimize drift in error rates of recalibrations and coherent errors, and (ii) maximize the stability and extend the duration of measurements [10-12].

To realize a practical quantum application, a quantum circuit constructed from a set of single-qubit, double-qubit, and/or multiple-qubit gates is synthesized into a QPU. In IBM terminologies, this synthesis process is termed the "transpilation" [3, 13]. The transpilation is the decomposition of a quantum circuit into a set of native gates. These native gates with their qubits are then routed, translated, and appropriately mapped into the architecture (the layout) of a QPU. The layout of a QPU defines how its physical qubits are arranged and connected together as neighboring and non-neighboring qubits. The neighboring qubits are directly communicated using double-qubit native gates; however, the non-neighboring qubits require a set of SWAP gates to switch their positions (indices) with other qubits for direct communications. The layout of `ibm_brisbane` QPU has a maximum set of four neighboring physical qubits; such that, there is one physical qubit in the middle connected to three physical qubits.

In this paper, as a proof of concept for our work, the `ibm_brisbane` QPU was utilized to transpile different quantum circuits consisting of a set of non-native gates, such gates are decomposed to their equivalent set of native gates, and the qubits of these native gates are then routed (if required) and mapped to match the physical qubits in the layout of `ibm_brisbane` QPU. We noticed that the `ibm_brisbane` QPU transpiles a quantum circuit of a few non-native gates into a cost-expensive quantum circuit of many native gates! Such a cost expense comes from: (i) the inefficient decomposition of non-native gates to native gates, and (ii) the limited layout connectivity of this QPU.

For instance, to decompose a non-native Toffoli gate, a number of SWAP gates are added to connect two controls with one target together, and adding more SWAP gates dramatically increases the quantum cost and depth of the final transpiled quantum circuit. A number of SWAP gates are added due to the inefficient decomposition of this Toffoli gate, i.e., the target qubit is inefficiently mapped between the two control qubits for the layout of this QPU. In this paper, for a final transpiled quantum circuit, the quantum cost is the total number of native gates, and the depth is the longest critical path through all native gates. When the depth is increased, the decoherence time (T2) [14-16] of the utilized physical qubits increases as well, which results in incorrect measurements.

The goal of this paper is to introduce a new quantum library for realizing cost-effective $n$-bit native gates, by efficiently transpiling $n$-bit non-native gates into IBM QPUs. These cost-effective $n$-bit gates are: (i) geometrically designed using the Bloch sphere, (ii) entirely built from the Clifford+T gates, and (iii) efficiently mapped into the layout of an IBM QPU. For that, we term this quantum library the "Clifford+T-based Architecture of Layout-Aware $n$-bit gates (CALA-$n$)", as well as to distinguish this CALA-$n$ from our previous work entitled "Generic Architecture of Layout-Aware $n$-bit gates (GALA-$n$)" [3], which is totally built from generic gates (as the Clifford+T gates and non-Clifford+T gates).

After transpiling with `ibm_brisbane` QPU, the $n$-bit gates of CALA-$n$ always have lower quantum costs and fewer depths than that of their equivalent transpiled non-native $n$-bit gates, due to the fact that SWAP gates are never added to the final transpiled quantum circuits when using CALA-$n$. In other words, for CALA-$n$, the target qubit of an $n$-bit gate is always mapped in the middle between the control qubits.

Due to the limited layout connectivity of IBM QPUs, CALA-$n$ has a set of efficiently designed $n$-bit gates, where $2 \leq n \leq 5$ qubits. These $n$-bit gates are: AND, NAND, OR, NOR, implication, inhibition [17, 18], controlled-$\sqrt{X}$, controlled-$\sqrt{X}^\dagger$, SWAP, and Fredkin. However, when the number of physical neighboring qubits of IBM QPUs is increased, the $n$ in CALA-$n$ increases as well. In other words, when the limited layout connectivity is decreased, the $n$ increases for greater than 5 qubits.

To comply with the limited layout connectivity of an IBM QPU, our 3-bit AND gate is mainly utilized to efficiently construct other $n$-bit gates of CALA-$n$, since we proved that this 3-bit AND gate is the cost-effective counterpart of the cost-expensive Toffoli gate. The Toffoli gate is mostly associated with building different quantum gates, e.g., $n$-bit controlled-$\sqrt{X}$, $n$-bit controlled-$\sqrt{X}^\dagger$, $n$-bit Fredkin, and $n$-bit Toffoli gates, where $n > 3$ qubits, for that various decompositions were discussed and proposed in [4, 19-23]. However, after transpiling different decompositions of the Toffoli gate, we observed that they always have higher quantum costs and deeper depths than that of our 3-bit AND gate of CALA-$n$. Concluding that our proposed CALA-$n$ can be efficiently utilized as a transpilation software package for IBM QPUs, to cost-effectively transpile $n$-bit non-native (controlled-$\sqrt{X}$, controlled-$\sqrt{X}^\dagger$, SWAP, Fredkin, and Toffoli) gates, in the context of lower quantum costs and fewer depths.

## 2 METHODOLOGY

The Clifford gates (or Clifford group) are a special group of single-qubit and double-qubit gates consisting of: I, X, Y, Z, H, $\sqrt{X}$, $\sqrt{X}^\dagger$, S, S$^\dagger$, CX, CY, CZ, and SWAP gates. These Clifford gates play a significant role in quantum computing, since they (i) invert the phase of a Pauli gate, and (ii) construct Pauli gates from other Pauli gates, as expressed in Eq. (1) below and stated in Table 1, where C is a Clifford gate, $P_\sigma$ is a Pauli gate, $P_\Sigma$ is the resultant Pauli gate, the symbol (†) is the conjugate transpose of a gate, and the symbol (·) is the matrix multiplication operator [4].

$$C \cdot P_\sigma \cdot C^\dagger = P_\Sigma \qquad (1)$$

**Table 1.** Transformations of Clifford gates for inverting the phases of Pauli gates and constructing other Pauli gates.

| C | · | $P_\sigma$ | · | $C^\dagger$ | = | $P_\Sigma$ |
|---|---|---|---|---|---|---|
| H | · | Z | · | H | = | X |
| S | · | X | · | S$^\dagger$ | = | Y |
| H | · | X | · | H | = | Z |
| Z | · | X | · | Z | = | $-$X |
| Z | · | Y | · | Z | = | $-$Y |
| X | · | Z | · | X | = | $-$Z |

Due to the importance of single-qubit T and T† gates and supporting them in most quantum systems [24-27], these gates are added to the Clifford gates to form the Clifford+T gates, for gaining more error-mitigation and fault-tolerant quantum computations on recent Noisy Intermediate-Scale Quantum (NISQ) devices [11, 16, 28]. In quantum computing, the Clifford+T gates can be either (i) expressed mathematically using the unitary matrices, as the numerical representation of their quantum operations, or (ii) represented geometrically using the Bloch sphere, as the visual representation of transitions for the states of a qubit after a gate (or gates) applies on it.

The Bloch sphere is a three-dimensional geometry of three axes (X-axis, Y-axis, and Z-axis), which visualizes the quantum operation (as an axial-rotated operation in Hilbert space $\mathcal{H}$) of a gate (or gates) applied on a qubit [1, 2]. Since the quantum operations of all IBM native gates are mainly rotating around the X-axis and Z-axis of the Bloch sphere, we propose the XY-plane (as the top-view of the Bloch sphere) to design the cost-effective $n$-bit gates of CALA-$n$, where $2 \leq n \leq 5$ qubits. In this XY-plane, an IBM (single-qubit or double-qubit) native gate rotates the state of a qubit by a counterclockwise rotational angle ($+\theta$) or a clockwise rotational angle ($-\theta$), in radians. Figure 1(a) demonstrates the Bloch sphere with its three axes and their corresponding angular rotations, and Figure 1(b) illustrates the XY-plane of the Bloch sphere, where no quantum operation is applied on a qubit yet.

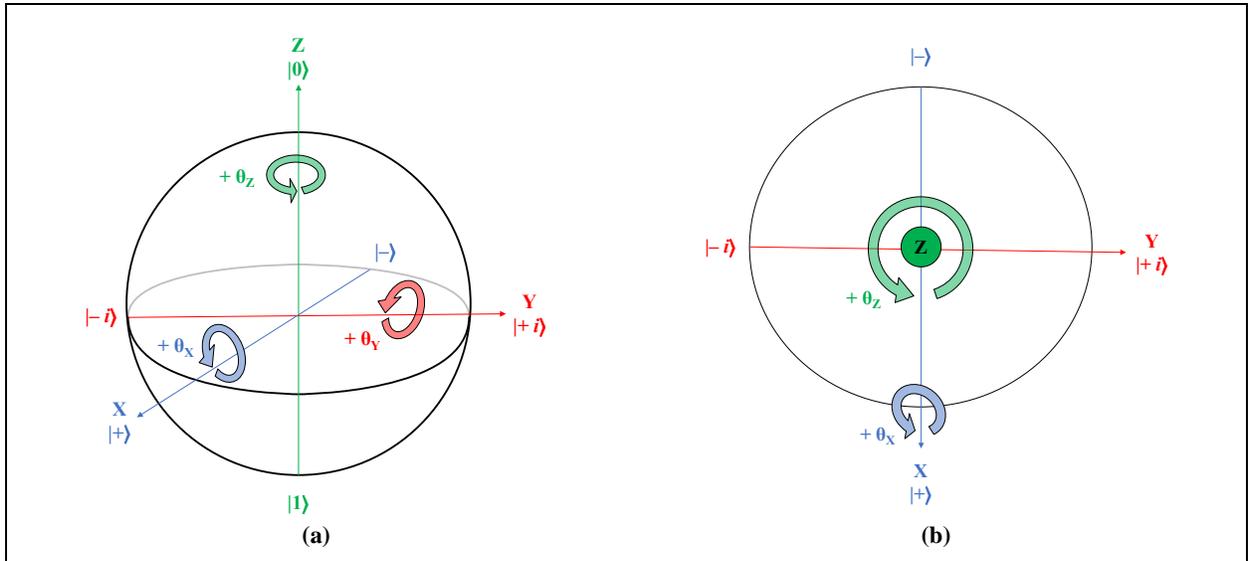

**Figure 1.** Schematics of the Bloch sphere consisting of: **(a)** the three axes (X-axis in blue, Y-axis in red, and Z-axis in green) with their corresponding counterclockwise rotational angles (+θx, +θy, and +θz), and **(b)** the XY-plane (as the top-view) for visualizing the quantum operations of IBM (single-qubit and double-qubit) native gates around the X-axis and Z-axis of the Bloch sphere.

In general, please observe that:

1. The Dirac notation [1, 2] is employed in this paper to represent a set of qubits for: (i) the transitions of states in the XY-plane of the Bloch sphere, and (ii) designing the $n$-bit gates of CALA-$n$, where the least significant qubit (LSQ) and the most significant qubit (MSQ) are on the right-side and the left-side of Dirac notation, respectively, e.g., |MSQ … LSQ⟩.

2. All IBM single-qubit and double-qubit native gates are unitary gates [1, 2], and note that a few of them are non-Hermitian gates [1, 2], i.e., their quantum operations are not their own inverses. For instance, the $\sqrt{X}^{\dagger}$ is the inverse gate of the $\sqrt{X}$ gate.
3. Some rotational gates alter the phase of a qubit, and the choice of rotational gates is a critical factor in designing a quantum circuit. For instance, the Z gate (as the π rotation around the Z-axis of the Bloch sphere) is not the same as the RZ(π) gate, due to the global phase difference, e.g., $RZ(\pi) = \begin{bmatrix} e^{-i\frac{\pi}{2}} & 0 \\ 0 & e^{i\frac{\pi}{2}} \end{bmatrix} = -iZ$ and the global phase is equal to '$-i$'.

In our work, we introduce the approach of using the Bloch sphere as a "geometrical design tool" to construct the $n$-bit gates of CALA-$n$, where $2 \leq n \leq 5$ physical qubits of an IBM QPU. To illustrate our Bloch sphere approach with ease, the XY-plane (as a circle) is divided into segments to represent the quantum rotations of Clifford+T gates as follows.

1. The "semicircle" segment is half of the XY-plane that represents the quantum rotations of Z gates. Such that, the XY-plane has two semicircles.
2. The "quadrant" segment is one-fourth of the XY-plane that represents the quantum rotations of S and $S^{\dagger}$ gates. Such that, the XY-plane has four quadrants.
3. The "octant" segment is one-eighth of the XY-plane that represents the quantum rotations of T and $T^{\dagger}$ gates. Such that, the XY-plane has eight octants.

For instance, Figure 2 illustrates these segments of the XY-plane, when a Clifford+T gate (Z, S, $S^{\dagger}$, T, or $T^{\dagger}$) is applied on the qubit of a gate, which is initially set to the |+⟩ state.

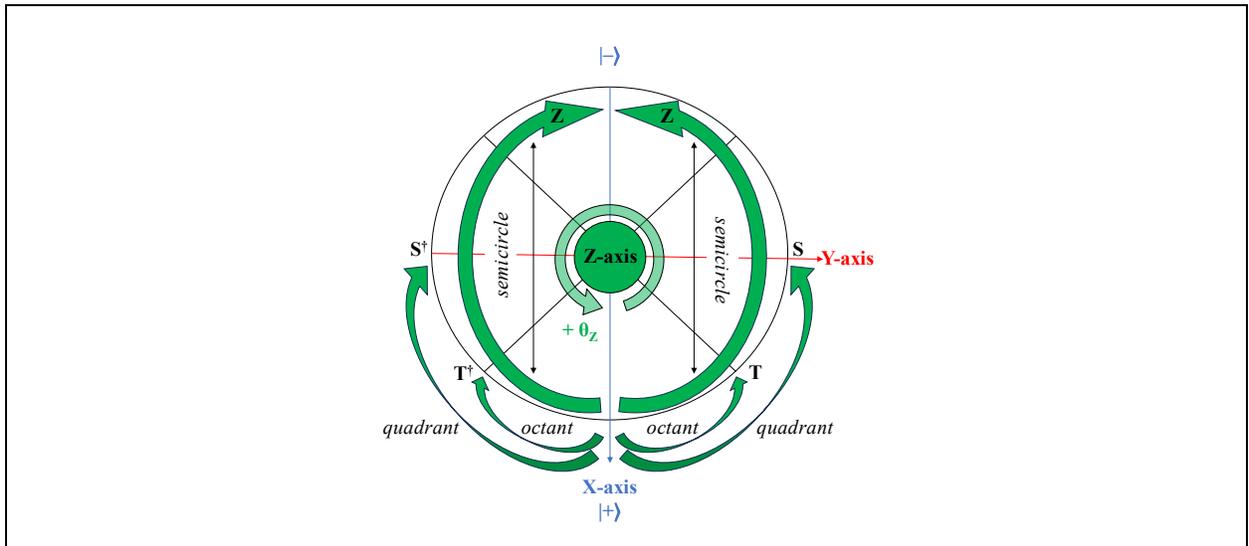

**Figure 2.** The segments of the XY-plane (as the top-view of the Bloch sphere) consisting of: (i) two semicircles for the quantum rotations of Z gates, (ii) four quadrants for the quantum rotations of S and $S^{\dagger}$ gates, and (iii) eight octants for the quantum rotations of T and $T^{\dagger}$ gates. The Z, S, $S^{\dagger}$, T, and $T^{\dagger}$ are a subset of the Clifford+T gates, which are applied on the qubit of a gate by rotating it around the center of this plane. Note that this qubit is initially set to the |+⟩ state.

Our proposed Bloch sphere approach mainly depends on the state transitions of a qubit in the XY-plane, which consists of the above-mentioned segments. To design a cost-effective gate of CALA-$n$ using the Bloch sphere approach, please observe our proposed "geometrical design steps" as follows.

**Step 1:** The qubit of a gate (as the target qubit) should be first placed on the XY-plane using the Clifford+T (H, $\sqrt{X}$, or $\sqrt{X}^\dagger$) gate, as a superposition gate.

**Step 2:** The Clifford+T (Z, S, $S^\dagger$, T, and $T^\dagger$) gates rotate the target qubit around the Z-axis of the Bloch sphere, i.e., they rotate a qubit around the center of the XY-plane.

**Step 3:** The Clifford+T (X) gate flips the position of the target qubit on the XY-plane around the X-axis of the Bloch sphere, i.e., the X gate flips a qubit on the opposite perimeter of the XY-plane in $\pi$ radians.

**Step 4:** The Clifford+T (CX) gate similarly acts as the X gate on the target qubit. However, the CX gate only operates when the other qubits (as the control qubits) of a gate are set to the $|1\rangle$ states. Note that the control qubits are not placed on the XY-plane, and they remain on the Z-axis of the Bloch sphere in $|0\rangle$ or $|1\rangle$ state.

**Step 5:** Based on the states of control qubits for a gate, the target qubit rotates around and/or flips in the XY-plane to reach its final state.

**Step 6:** Finally, another Clifford+T (H, $\sqrt{X}$, or $\sqrt{X}^\dagger$) gate re-places the final reached state of the target qubit from the XY-plane to the Z-axis of the Bloch sphere. Such a replacement is considered the final output of a gate, for either the $|1\rangle$ state (as a solution) or the $|0\rangle$ state (as a non-solution).

From the geometrical design steps, when the target qubit of a gate initially sets to the $|0\rangle$ state, the choice of a Clifford+T (H, $\sqrt{X}$, or $\sqrt{X}^\dagger$) gate (in Step 1) plays a crucial role in the Bloch sphere approach, since such a superposition gate determines where the target qubit should be placed in the XY-plane. Hence, the subsequent steps (from Step 2 to Step 6) should be carefully taken into account depending on the chosen superposition gate as follows.

1. The H gate places the target qubit on the XY-plane at the $|+\rangle$ state.
2. The $\sqrt{X}$ gate places the target qubit on the XY-plane at the $|-i\rangle$ state.
3. The $\sqrt{X}^\dagger$ gate places the target qubit on the XY-plane at the $|+i\rangle$ state.

In this paper, we design various cost-effective $n$-bit gates of CALA-$n$ for IBM QPUs, and these $n$-bit gates of CALA-$n$ are: visually designed using our proposed Bloch sphere approach, entirely constructed using Clifford+T gates only, and efficiently mapped to well-suit the layout of an IBM QPU. Table 2 states these cost-effective $n$-bit gates of CALA-$n$, based on their quantum purposes as the square-rooting, switching, Boolean, and distance gates, where $2 \leq n \leq 5$ qubits.

**Table 2.** Summary of cost-effective *n*-bit gates of CALA-*n* based on their quantum purposes (square-rooting, switching, Boolean, and distance), where $2 \leq n \leq 5$ qubits.

| | *n*-bit gates of CALA-*n* | | | |
|---|---|---|---|---|
| *n* qubits | Square-rooting gates | Switching gates | Boolean gates | Distance gates |
| 2 | controlled-$\sqrt{X}$ and controlled-$\sqrt{X}^{\dagger}$ | SWAP | – | – |
| 3 | | Fredkin | AND, NAND, OR, NOR, implication, and inhibition | Miller [29-31] |
| 4 | – | | | – |
| 5 | – | – | | – |

## 2.1 The 2-bit gates of CALA-*n*

The 2-bit gates of controlled-$\sqrt{X}$, controlled-$\sqrt{X}^{\dagger}$, and SWAP play important roles in quantum computing, since they are mainly used to build Peres gate [31, 32], inverse Peres gate [32], arbitrary gates for the majority of three variables in a function [29, 30, 32-35], just to name a few. These three 2-bit gates are IBM non-native double-qubit gates. Therefore, when these three 2-bit gates are transpiled with an IBM QPU, their final transpiled quantum circuits will have higher quantum costs and deeper depths, as shown in Figure 3. For these reasons, we re-design these three 2-bit gates using the Bloch sphere approach, to form a new set of cost-effective 2-bit gates for IBM QPUs with lower quantum costs and fewer depths, as shown in Figure 4 and stated in Table 3.

From the Bloch sphere approach, the 2-bit gates of controlled-$\sqrt{X}$ and controlled-$\sqrt{X}^{\dagger}$ are simply re-designed using only one CX gate, with a fewer number of single-qubit gates. However, the SWAP gate is directly re-designed based on the iSWAP gate proposed by the IBM quantum platform [36].

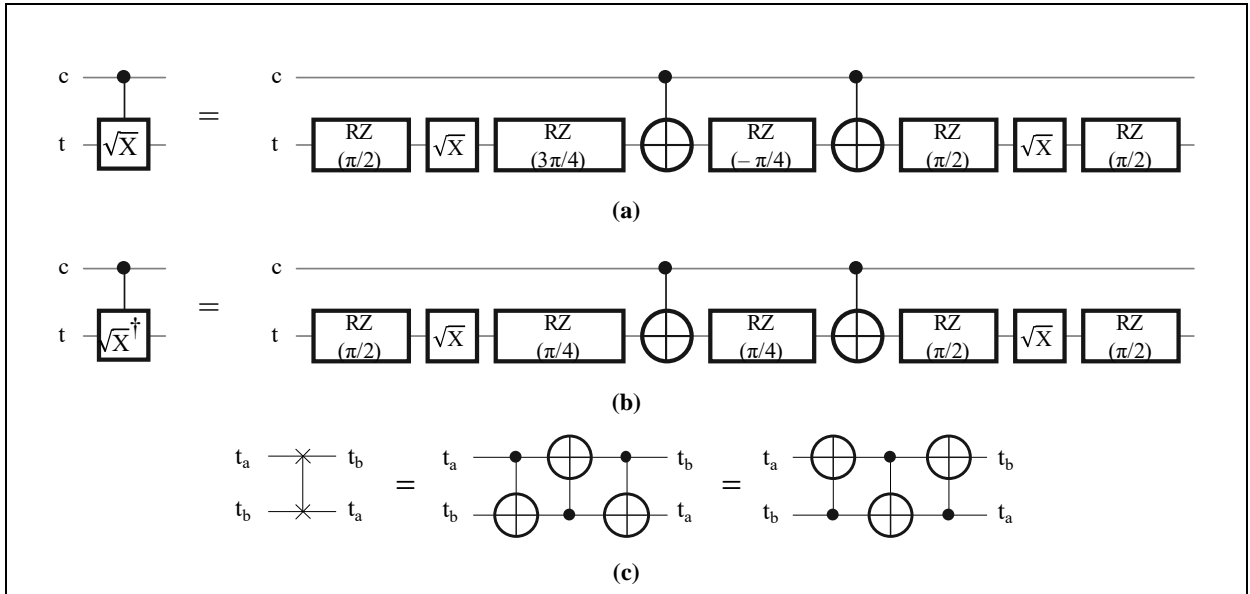

**Figure 3.** An IBM QPU standard transpilation from non-native gates to native gates for the 2-bit gates of: **(a)** controlled-$\sqrt{X}$ gate, **(b)** controlled-$\sqrt{X}^{\dagger}$ gate, and **(c)** SWAP gate, where a and b are the indices of two qubits, c is a control qubit, and t is a target qubit.

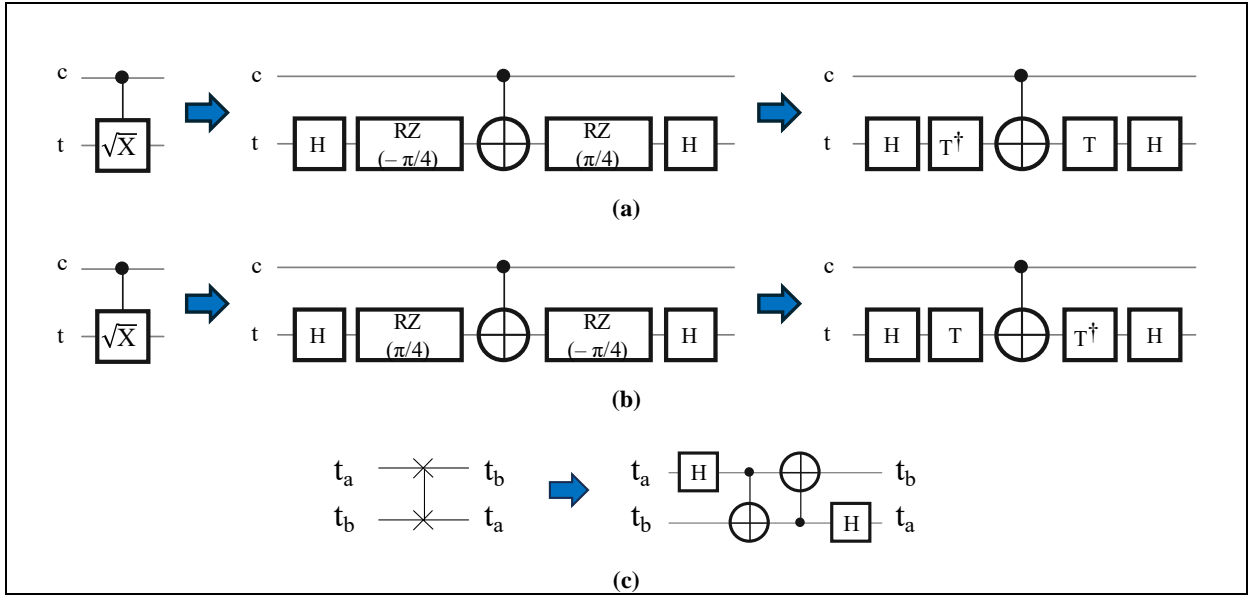

**Figure 4.** Re-designing IBM non-native gates to IBM native gates using the Bloch sphere approach for the 2-bit gates of: **(a)** controlled-√X gate, **(b)** controlled-√X† gate, and **(c)** SWAP gate, where a and b are the indices of two qubits, c is a control qubit, and t is a target qubit. Note that (i) the Bloch sphere approach only utilizes Clifford+T gates, (ii) the SWAP gate is re-designed based on the iSWAP gate proposed by the IBM quantum platform [36], and (iii) the H gate is an IBM non-native gate, so further decomposition is required to the sequence of three IBM native gates {S √X S} = {RZ(π/2) √X RZ(π/2)}.

**Table 3.** Comparison of the quantum costs and depths between the IBM standard transpilation and the Bloch sphere approach for the 2-bit gates (controlled-√X, controlled-√X†, and SWAP), based on the final counts of IBM single-qubit (√X, X, and RZ) and double-qubit (CX) native gates for the final transpiled quantum circuits of these 2-bit gates.

| | Counts of IBM native gates and depths for the final transpiled quantum circuits | | | | | | | | | |
|---|---|---|---|---|---|---|---|---|---|---|
| | **Using IBM standard transpilation** | | | | | **Using the Bloch sphere approach** | | | | |
| **2-bit gates** | **√X** | **X** | **CX** | **RZ** | **Depth** | **√X** | **X** | **CX** | **RZ** | **Depth** |
| **controlled-√X** | 2 | – | 2 | 5 | 9 | 2 | – | 1 | 4 | 7 |
| **controlled-√X†** | 2 | – | 2 | 5 | 9 | 2 | – | 1 | 4 | 7 |
| **SWAP** | – | – | 3 | – | 3 | 2 | – | 2 | 4 | 3 |

## 2.2 The 3-bit AND gate of CALA-*n*

In this paper, we geometrically prove that the quantum operation of the 3-bit AND gate is the cost-effective counterpart of the Toffoli gate. Therefore, the 3-bit AND gate can be utilized to construct various *n*-bit gates of CALA-*n*, where $3 \leq n \leq 5$ qubits. The 3-bit AND gate consists of three qubits: two control qubits and one target qubit. By using the Bloch sphere approach, the 3-bit AND gate is designed using the geometrical design steps, based on the following quantum design constraints.

1. *The XY-plane of the Bloch sphere*: Two H gates are utilized here, the first H gate places the target qubit (initially set to the $|0\rangle$ state) to the XY-plane, while the second H gate returns this qubit to the Z-axis of the Bloch sphere, as a $|1\rangle$ state (solution) or a $|0\rangle$ state (non-solution).
2. *The Clifford+T gates*: The two controls flip the target qubit (in the XY-plane) based on their states, using a set of CX gates. The T and $T^\dagger$ gates rotate the target qubit around the center of the XY-plane.
3. *The limited layout connectivity of an IBM QPU*: The target qubit is efficiently mapped into the middle between the two control qubits, so SWAP gates are never added to the final transpiled quantum circuit.

The 3-bit AND gate is designed as a quantum circuit with a symmetrical structure, as originally introduced by Barenco *et al.* [4] for decomposing the Toffoli gate as shown in Figure 5(a). Note that Barenco *et al.* utilized the unitary matrices and RY gates for such a decomposition. The RY gates do not belong to the Clifford+T gates, as well as they are IBM non-native gates, so further decompositions are required. However, we generalize such a symmetrical structure using the Bloch sphere approach to: (i) select the appropriate set of Clifford+T gates for the 3-bit AND gate, and (ii) develop various *n*-bit gates of CALA-*n* based on this 3-bit AND gate. Table 4 states the differences between the work of Barenco *et al.* and our work in this paper.

For the symmetrical structure of the 3-bit AND gate, the identical neighboring Clifford+T gates will cancel each other, when any of the controls is in the $|0\rangle$ state (as a false Boolean value), and then the target qubit remains at its initial $|0\rangle$ state (as the indication of a non-solution). For this reason, we are interested here in decomposing and generalizing the quantum gates with symmetrical structures, and not with the non-symmetrical structures, which are difficult to be generalized using the Bloch sphere approach based on the layout of an IBM QPU, i.e., by directly connecting all control qubits to the target qubit and without any connection between the control qubits. For instance, Figure 5(b) demonstrates how the IBM quantum system transpiles a Toffoli gate (two controls and one target) into a quantum circuit with a non-symmetrical structure, where all control and target qubits are connected to each other. Note that the connection between one qubit to another qubit means that these two qubits utilize one CX gate; thus, one qubit represents a control while another qubit represents the target.

**Table 4.** Decomposition questionnaires and differences between the work of Barenco *et al.* and our work.

| Questions | Barenco *et al.* [4] (Toffoli gate) | Our work (3-bit AND gate of CALA-*n*) |
|---|---|---|
| **Has a symmetrical structure?** | Yes | Yes |
| **How are the gates chosen?** | Unitary matrices and mathematical calculations | Bloch sphere approach and Clifford+T gates |
| **If the Bloch sphere is used, which plane is utilized?** | The Bloch sphere is not used | The Bloch sphere is used: The XY-plane |
| **Which Boolean operations can be derived?** | AND | AND, NAND, OR, NOR, and implication, inhibition |
| **Suitable for IBM QPUs?** | No, RY and CZ gates need to be decomposed (expensive circuits) | Yes, decompositions are not required (cheap circuits based on quantum layouts) |

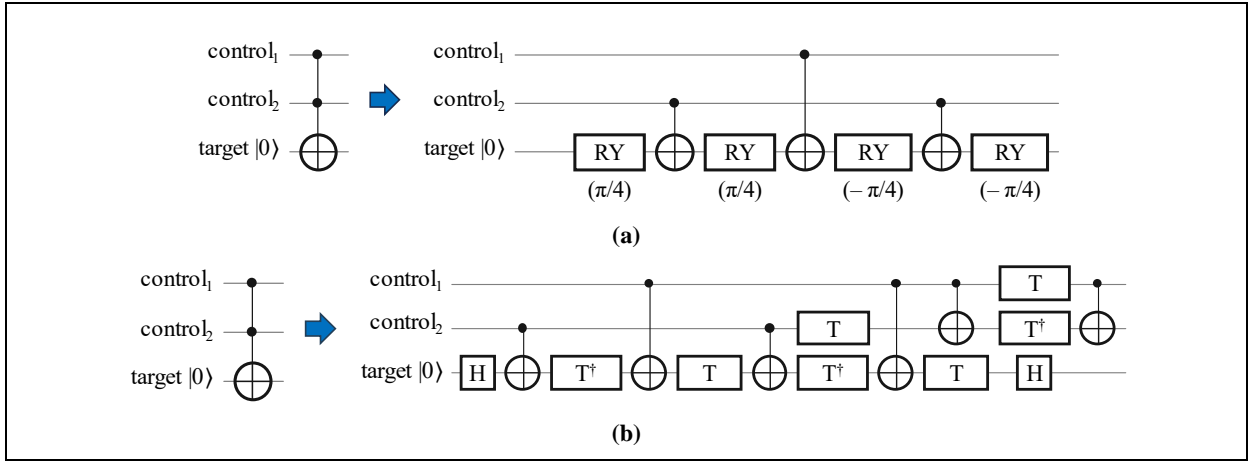

**Figure 5.** Decompositions of a Toffoli gate into the quantum circuits with: **(a)** a symmetrical structure [4] for all control qubits connected to the target qubit without any connection between the controls, and **(b)** a non-symmetrical structure for all qubits (controls and target) connected to each other, which is the outcome of IBM QPU transpilation. Note that the RY gates do not belong to the Clifford+T gates, and they are IBM non-native gates.

Figure 6 illustrates our generalized symmetrical structure for the quantum circuit of 3-bit AND gate of CALA-$n$, which is composed of three Clifford+T double-qubit gates (as CX) and the following Clifford+T single-qubit gates.

1. The $SP_1$ and $SP_2$ are the superposition (H, $\sqrt{X}$, and/or $\sqrt{X}^\dagger$) gates.
2. The $AX_1$ and $AX_2$ are the auxiliary gates. Both $AX_1$ and $AX_2$ will be utilized later for constructing various $n$-bit gates of CALA-$n$, and $AX_1 = AX_2 = I$ for the 3-bit AND gate.
3. $\theta$ (as $\theta_1$, $\theta_2$, $\theta_3$, and $\theta_4$) denotes four Clifford+T gates. Due to the symmetrical structure of the 3-bit AND gate, we assume that $\theta_1 = \theta_3$ and $\theta_2 = \theta_4$.

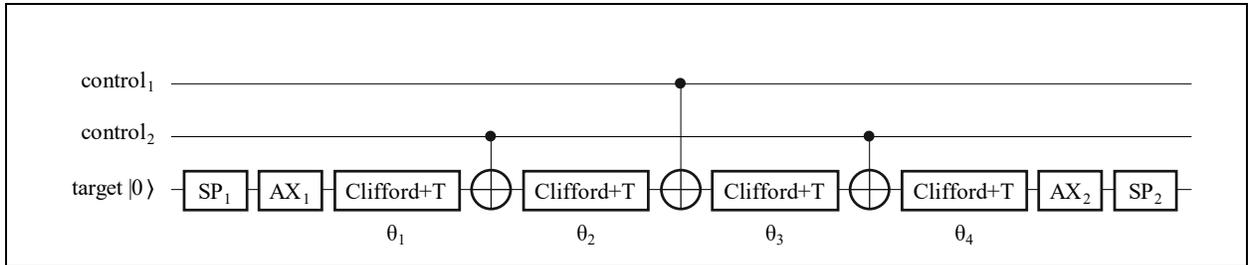

**Figure 6.** Our generalized symmetrical structure for the quantum circuit of 3-bit AND gate of CALA-$n$, where $\theta$ ($\theta_1$, $\theta_2$, $\theta_3$, $\theta_4$) corresponds to four Clifford+T gates ($\theta_1 = \theta_3$ and $\theta_2 = \theta_4$), $SP_1$ and $SP_2$ are the superposition gates, $AX_1$ and $AX_2$ are the auxiliary gates for constructing various $n$-bit gates of CALA-$n$ ($3 \leq n \leq 5$), and the target qubit is initially set to the $|0\rangle$ state.

In Figure 6, all control qubits are directly connected to the target qubit using three CX gates, without any connection between the controls. Such unidirectional connections guarantee two advantages: (i) the target qubit can be switched with any of the control qubits, to match the quantum layout of an IBM QPU, i.e., the target qubit is placed in the middle between the two control qubits, and (ii) the SWAP gates will not be added to the final transpiled quantum circuit of this generalized symmetrical structure, i.e., this yields to have a lower quantum cost and a fewer depth.

The superposition gates (SP$_1$ and SP$_2$) are defined here as H gates, to place the target qubit from/to the Z-axis and the XY-plane of the Bloch sphere. To provide our proposed CALA-$n$ as an open and generic quantum library, other superposition gates can be utilized for SP$_1$ and SP$_2$. For instance, the $\sqrt{X}$ and $\sqrt{X}^\dagger$ gates can be utilized as the superposition gates for SP$_1$ and SP$_2$ gates in different permutations. However, different geometrical analyses should be taken into account when choosing these $\sqrt{X}$ and $\sqrt{X}^\dagger$ gates for the 3-bit AND gate. Note that the $\sqrt{X}$ gate is an IBM native gate, while the H and $\sqrt{X}^\dagger$ gates are IBM non-native gates, for that further decompositions are required.

From Figure 2 and Figure 6, the Clifford+T gates (as $\theta_1$, $\theta_2$, $\theta_3$, and $\theta_4$) of the 3-bit AND gate are accurately chosen based on: (i) the symmetrical structure of this gate, (ii) the segmental analyses of two semicircles, four quadrants, and eight octants, and (iii) the flipping functionalities of three CX gates. Subsequently, based on this symmetrical structure and the XY-plane of the Bloch sphere, the following "geometrical design rules" for the 3-bit AND gate choose the accurate Clifford+T gates ($\theta_1$, $\theta_2$, $\theta_3$, and $\theta_4$), and these rules are step-by-step restricting possible choices of segments and Clifford+T gates.

**Rule 1:** Define all Clifford+T gates as a set: CTG$_0$ = {I, X, Y, Z, H, $\sqrt{X}$, $\sqrt{X}^\dagger$, S, S$^\dagger$, T, T$^\dagger$, CX, CY, CZ, SWAP}.

**Rule 2:** Define all segments of the XY-plane as a set: SEG$_0$ = {semicircles, quadrants, octants}.

**Rule 3:** Since the target qubit is not controlling other qubits, then CTG$_0$ is limited to a set of single-qubit Clifford+T gates: CTG$_1$ = {I, X, Y, Z, H, $\sqrt{X}$, $\sqrt{X}^\dagger$, S, S$^\dagger$, T, T$^\dagger$}.

**Rule 4:** Since SEG$_0$ only identifies Clifford+T gates rotating by the Z-axis of the Bloch sphere, then CTG$_1$ is limited to a set of single-qubit Z-based gates: CTG$_2$ = {Z, S, S$^\dagger$, T, T$^\dagger$}.

**Rule 5:** Since there are four θ (as Clifford+T gates in Figure 6) for $\theta_1 = \theta_3$ and $\theta_2 = \theta_4$, then the semicircles (as Z gates) are excluded to prevent similar repetitive circular rotations in the XY-plane; such that, SEG$_1$ is restrictive of SEG$_0$ and CTG3 is restrictive of CTG$_2$: SEG$_1$ = {quadrants, octants} and CTG$_3$ = {S, S$^\dagger$, T, T$^\dagger$}.

**Rule 6:** Based on Rule 5, the four Clifford+T gates are reduced to two gates, and the quadrants are excluded to prevent similar repetitive circular rotations in the XY-plane: SEG$_2$ = {octants} and CTG$_4$ = {T, T$^\dagger$}.

**Rule 7:** Apply CTG$_4$ (for all permutations of T and T$^\dagger$) with three CX gates (for all states of control qubits) on SEG$_2$ (octants) of the XY-plane, then the correct permutations of T and T$^\dagger$ that implement the Boolean AND function are plugged into the four Clifford+T gates (as $\theta_1$, $\theta_2$, $\theta_3$, and $\theta_4$) of 3-bit AND gate.

For SP$_1$=SP$_2$=H and AX$_1$=AX$_2$=I, after applying the above-mentioned geometrical design rules to construct the 3-bit AND gate, i.e., to implement the Boolean AND function, the suitable values for the four Clifford+T gates in the sequence of {$\theta_1$, $\theta_2$, $\theta_3$, $\theta_4$} are generated in the sequence of {T$^\dagger$, T, T$^\dagger$, T} (or {T, T$^\dagger$, T, T$^\dagger$}), respectively. Table 5 analytically verifies these geometrical design rules for all states of control qubits, and Figure 7 visually illustrates these verifications. Hence, the 3-bit AND gate is geometrically designed and proven as the functional counterpart of the Toffoli gate. Figure 8 shows the complete quantum circuit of the 3-bit AND gate of CALA-$n$, which is completely constructed using a restricted set of Clifford+T gates (I, H, T, T$^\dagger$, and CX).

**Table 5.** Analytical verifications for the accurate four Clifford+T gates ($\theta_1$, $\theta_2$, $\theta_3$, and $\theta_4$) of the 3-bit AND gate, when the target qubit is initially set to the $|0\rangle$ state and $AX_1=AX_2=I$. Note that $\frac{\alpha \pi}{\beta} = \frac{1}{\sqrt{2}}\left(|0\rangle + e^{i\frac{\alpha\pi}{\beta}}|1\rangle\right)$, $\alpha$ and $\beta$ are numerical values, and $e^{\pm ix}$ is the Euler's formula of $\cos(x) \pm i\sin(x)$.

| | **Applying quantum gates on the target qubit** | | | | | | | | | |
|---|---|---|---|---|---|---|---|---|---|---|
| $\|control_2\,control_1\rangle$ | $SP_1$ (H) | $\theta_1$ ($T^\dagger$) | CX of $control_2$ | $\theta_2$ (T) | CX of $control_1$ | $\theta_3$ ($T^\dagger$) | CX of $control_2$ | $\theta_4$ (T) | $SP_2$ (H) | As Boolean values |
| $\|0\,0\rangle$ | $\|+\rangle$ | $\frac{7\pi}{4}$ | – | $\|+\rangle$ | – | $\frac{7\pi}{4}$ | – | $\|+\rangle$ | $\|0\rangle$ | False |
| $\|0\,1\rangle$ | $\|+\rangle$ | $\frac{7\pi}{4}$ | – | $\|+\rangle$ | $\|+\rangle$ | $\frac{7\pi}{4}$ | – | $\|+\rangle$ | $\|0\rangle$ | False |
| $\|1\,0\rangle$ | $\|+\rangle$ | $\frac{7\pi}{4}$ | $\frac{\pi}{4}$ | $\|+i\rangle$ | – | $\frac{\pi}{4}$ | $\frac{7\pi}{4}$ | $\|+\rangle$ | $\|0\rangle$ | False |
| $\|1\,1\rangle$ | $\|+\rangle$ | $\frac{7\pi}{4}$ | $\frac{\pi}{4}$ | $\|+i\rangle$ | $\|-i\rangle$ | $\frac{5\pi}{4}$ | $\frac{3\pi}{4}$ | $\|-\rangle$ | $\|1\rangle$ | **True** |

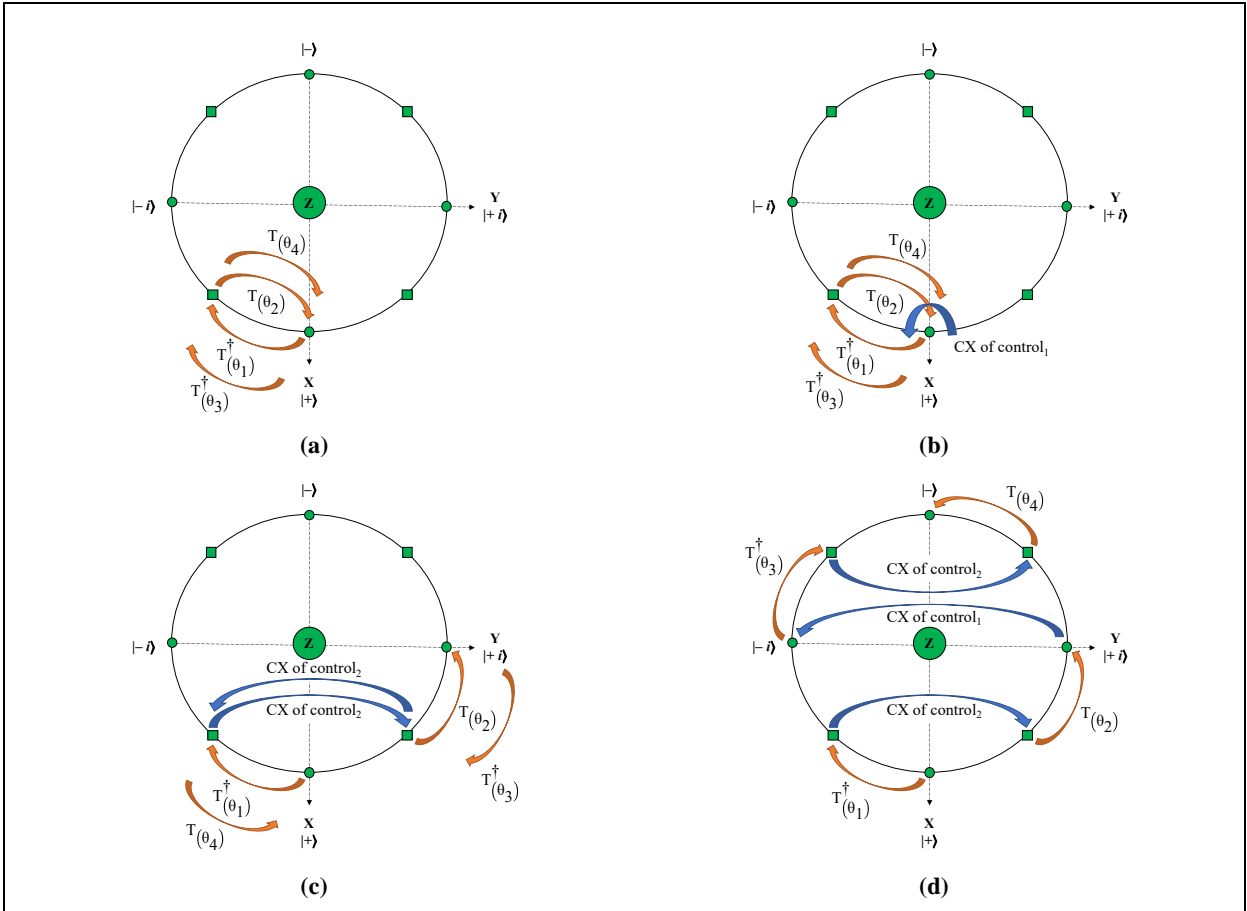

**Figure 7.** Schematics of the XY-plane of the Bloch sphere for verifying the accurate four Clifford+T gates ($\theta_1=T^\dagger$, $\theta_2=T$, $\theta_3=T^\dagger$, and $\theta_4=T$) for the 3-bit AND gate of CALA-$n$, for all states of control qubits ($control_1$ and $control_2$): **(a)** both controls = $|0\rangle$, **(b)** $control_1 = |1\rangle$ and $control_2 = |0\rangle$, **(c)** $control_1 = |0\rangle$ and $control_2 = |1\rangle$, and **(d)** both controls = $|1\rangle$, where the target qubit initially sets to the $|0\rangle$ state, the arrows in orange indicate the octant-based rotations around the Z-axis of the Bloch sphere, and the arrows in blue indicate the rotations around the X-axis of the Bloch sphere.

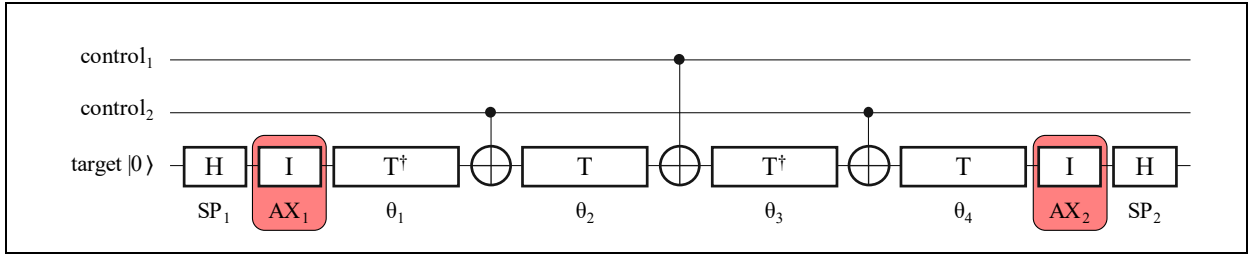

**Figure 8.** The generalized symmetrical structure for the quantum circuit of 3-bit AND gate of CALA-*n* with the accurately verified four Clifford+T gates ($\theta_1=\theta_3=T^\dagger$ and $\theta_2=\theta_4=T$), when $SP_1=SP_2=H$ and $AX_1=AX_2=I$ and. Note that the $AX_1$ and $AX_2$ gates can be removed from this structure (as indicated in red areas), to further reduce the quantum cost and the depth as well.

## 2.3 The 3-bit Boolean gates of CALA-*n*

From Figure 6 and Figure 8, when the four Clifford+T gates ($\theta_1$, $\theta_2$, $\theta_3$, and $\theta_4$) are not symmetrical anymore, various 3-bit gates of CALA-*n* can be constructed using the Bloch sphere approach. By repeating the above-mentioned geometrical design rules for these four non-symmetrical Clifford+T gates, the restrictive sets of $CTG_4 = \{S, S^\dagger, T, T^\dagger\}$ and $SEG_2$ = {quadrants, octants} are then obtained, and the minimum number of for such 3-bit gates of CALA-*n* is 256 gates, by assuming that $SP_1=SP_2=H$ and $AX_1=AX_2=I$, as stated in Eq. (2), where $\|.\|$ indicates the total number of gates and $\theta = \{\theta_1, \theta_2, \theta_3, \theta_4\}$.

$$\text{Number of 3-bit gates in CALA-}n = \|SP_1\| \times \|AX_1\| \times \| CTG_4 \|^{\|\theta\|} \times \|AX_2\| \times \|SP_2\| \qquad (2)$$

From Eq. (2), when the superposition and auxiliary gates have all permutative sets for their respective Clifford+T single-qubit gates, i.e., $SP_1=SP_2=\{H, \sqrt{X}, \sqrt{X}^\dagger\}$ and $AX_1=AX_2=\{I, X, \sqrt{X}, \sqrt{X}^\dagger, Z, S, S^\dagger, T, T^\dagger\}$, the maximum number of all 3-bit gates of CALA-*n* is equal to 186624 gates! Hence, such gates can implement massive numbers of quantum operations, including Boolean gates synthesizing, quantum gates synthesizing, phase inversion, just to name a few.

However, in this paper, we are interested in constructing quantum Boolean-based gates, such as AND, NAND, OR, NOR, implication, and inhibition. These quantum Boolean-based gates can be used to effectively build Boolean oracles [37, 38], especially for the problems structured in Product-of-Sums (POS) [39], Sum-of-Products (SOP) [40], Exclusive-or Sum-of-Products (ESOP) [41-43], constraints satisfiable problems–satisfiability (CSP–SAT) [44], and in designing XOR functions of two product gates of different Hamming distances [43, 45], just to name a few.

This paper is an introductory research for building the fundamental cost-effective *n*-bit gates of CALA-*n*, and further geometrical analyses of the Bloch sphere can be utilized to construct various *n*-bit specific-task gates, such as AND-XOR SAT, OR-XOR SAT, majority, implication-inhibition, Phase-based gates for Phase oracles [37, 38], just to name a few. Hence, CALA-*n* is an open and generic quantum library for prospective quantum computing research, to develop interesting and cost-effective *n*-bit quantum gates, which can be easily adapted with different layouts of physical neighboring qubits for superconducting quantum systems, e.g., the QPUs of IBM, Google, and Rigetti.

Table 6 states six 3-bit Boolean gates of CALA-$n$ that we are interested in our research, which are: AND, NAND, OR, NOR, implication, and inhibition, where the target qubit is initially set to the $|0\rangle$ state, $SP_1=SP_2=H$, and $AX_1 = I$. Note that, in Table 6, (i) all 3-bit Boolean gates of CALA-$n$ are completely constructed using Clifford+T gates (I, H, Z, T, $T^\dagger$, and CX), (ii) the $AX_2$ of a 3-bit Boolean-inverted gate is added as a Z (or $-Z$) gate, to invert the solutions of its 3-bit Boolean gate, e.g., NAND and AND gates, (iii) the $-Z$ gate can be easily constructed using the Clifford transformations as stated in Table 1, and (iv) the Z and $-Z$ gates are accurately chosen to mitigate the phase differences in solutions, as illustrated in Figure 9 using the q-spheres [46]. In contrast to the Bloch sphere visualizing the states of one qubit, the q-sphere visualizes the states of multiple qubits, where (i) the size of a circle denotes the amplitude probability of a qubit, (ii) the color of a circle represents the global phase of a qubit, and (iii) the Dirac notations of all qubits are in the form of $|target\ control_2\ control_1\rangle$.

**Table 6.** Configurations of $SP_1$, $SP_2$, $AX_1$, $AX_2$, and θ (four Clifford+T gates) for six 3-bit Boolean gates of CALA-$n$, where the target qubit is initially set to $|0\rangle$ state, $a$ is the first control (control$_1$), and $b$ is the second control (control$_2$).

| 3-bit Boolean gates | Configurations of Clifford+T gates applied on the target qubit ||||||||
|---|---|---|---|---|---|---|---|---|
| | $SP_1$ | $AX_1$ | $\theta_1$ | $\theta_2$ | $\theta_3$ | $\theta_4$ | $AX_2$ | $SP_2$ |
| **AND** $(a \wedge b)$ | H | I | $T^\dagger$ | T | $T^\dagger$ | T | I | H |
| **NAND** $\neg (a \wedge b)$ | H | I | $T^\dagger$ | T | $T^\dagger$ | T | $-Z$ | H |
| **OR** $(a \vee b)$ | H | I | T | T | T | T | Z | H |
| **NOR** $\neg (a \vee b)$ | H | I | T | T | T | T | I | H |
| **Implication** $(\neg a \vee b)$ | H | I | $T^\dagger$ | $T^\dagger$ | T | T | $-Z$ | H |
| **Inhibition** $\neg (\neg a \vee b)$ | H | I | $T^\dagger$ | $T^\dagger$ | T | T | I | H |

From Table 6, all 3-bit Boolean gates of CALA-$n$ have identical quantum circuits with symmetrical structures as previously shown in Figure 6, except for θ ($\theta_1$, $\theta_2$, $\theta_3$, $\theta_4$) and $AX_2$ gates. Hence, this gives an intuition to investigate various Boolean gates and operators, as they will be discussed in the next subsection.

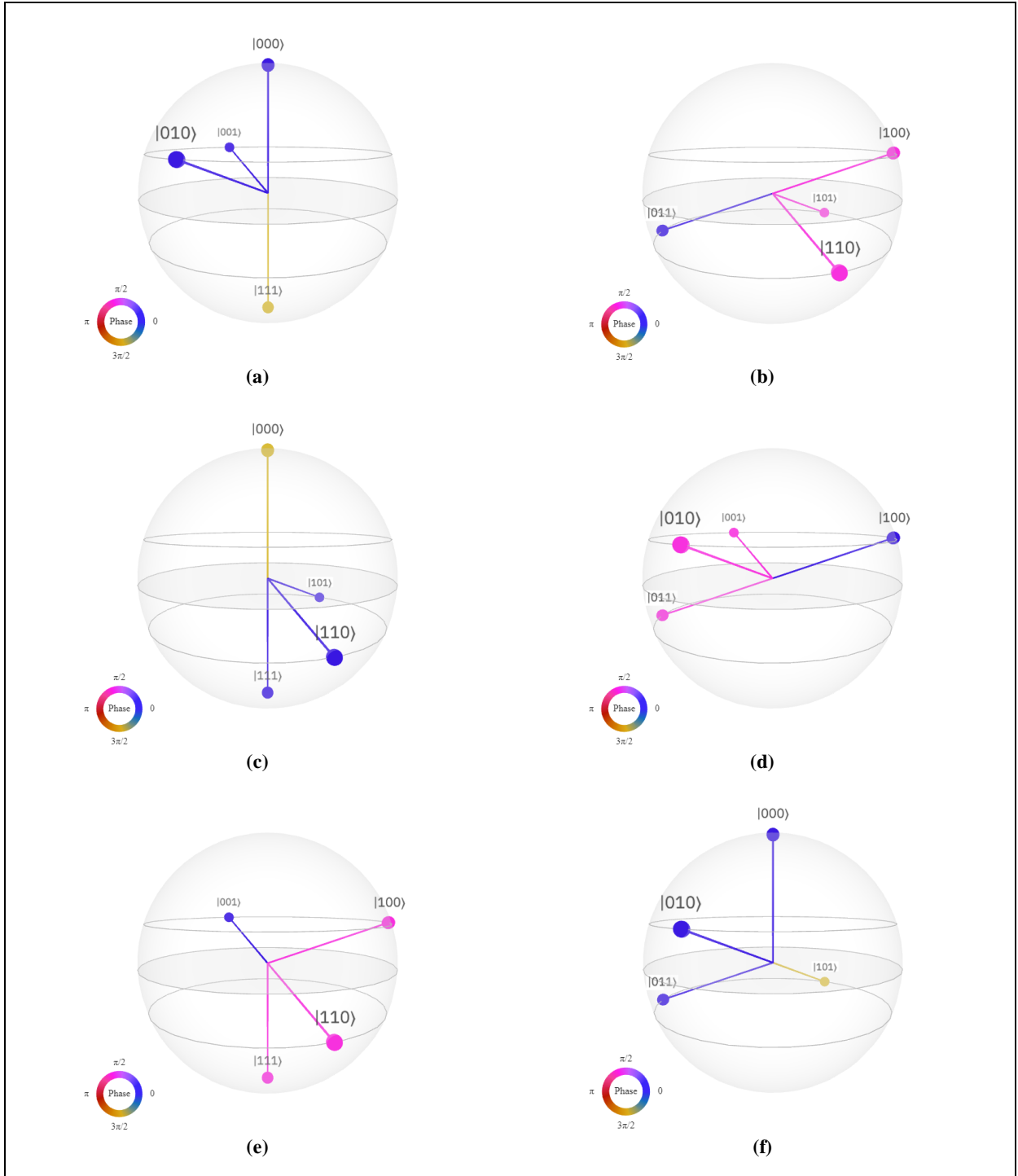

**Figure 9.** Q-spheres [46] for the outcomes of 3-bit Boolean gates of CALA-$n$ (see Table 6): **(a)** AND, **(b)** NAND, **(c)** OR, **(d)** NOR, **(e)** implication, and **(f)** inhibition. The Dirac notations for all qubits are in the form of |target control$_2$ control$_1$⟩, where all target qubits are initially set to the |0⟩ state. For the global phases, the blue indicates the 0 radians, the pink indicates the $\pi/2$ radians, and the yellow indicates the $3\pi/2$ radians.

## 2.4 The *n*-bit gates of CALA-*n*

The 3-bit AND gate is mainly used to hierarchically construct various *n*-bit gates of CALA-*n*, where *n* is restricted to the range (3, 4, 5) of qubits. This range restriction is related to the limited layout connectivity for the physical neighboring qubits of IBM QPUs. In this paper, we choose the `ibm_brisbane` QPU as an example of such a limited layout connectivity. Note that the *n*-bit gates of CALA-*n* consist of *n*–1 control qubits and one target qubit. In contrast, the *n*-bit Fredkin gates always have *n*–2 control qubits and two target qubits. In this paper, the constructed *n*-bit gates of CALA-*n* are categorized as follows (see Table 2).

1. *n-bit Boolean gates*: AND, NAND, OR, NOR, implication, and inhibition gates, where $3 \leq n \leq 5$ qubits.
2. *n-bit switching gates*: Fredkin gates, where $3 \leq n \leq 4$ qubits.
3. *3-bit square-rooting gates*: controlled-$\sqrt{X}$ and controlled-$\sqrt{X}^{\dagger}$ gates, and these gates are built using the 2-bit gates of controlled-$\sqrt{X}$ and controlled-$\sqrt{X}^{\dagger}$, as shown in Figure 4(a) and Figure 4(b), respectively.
4. *3-bit distance gate*: Miller gate.

Please observe that to overcome the limited layout connectivity of an IBM QPU, an *m* number of ancilla qubits is utilized when constructing the above-mentioned *n*-bit gates of CALA-*n*, and the *m* number of ancilla qubits is not counted with the number of *n* qubits in CALA-*n*, where $m \geq 0$ qubits.

The powers of CALA-*n* as a generic quantum library are unlimited for constructing various configurable gates, operators, and structures. Such that, all 3-bit Boolean gates of CALA-*n* are capable of constructing different *n*-bit gates in various configurations. For instance, a 5-bit OR-AND-OR gate (as a POS structure), a 5-bit AND-OR-AND gate (as a SOP structure), a 5-bit implication-NAND-inhibition gate (as a customized Boolean structure), and so on. Figure 10 illustrates different *n*-bit gates and structures of CALA-*n*, where $3 \leq n \leq 5$. These different gates and structures are then cost-effectively mapped into the layout of `ibm_brisbane` QPU without utilizing any SWAP gates, even when utilizing *m* ancilla qubits as discussed in the next subsection.

## 2.5 Native Gates and Layouts of IBM QPUs

At the time of writing this paper, for all IBM QPUs of 127 qubits (including `ibm_brisbane`), the IBM quantum system replaces its native Feynman (CX) gate with the ECR (echoed cross-resonance) gate [10-12] to: (i) minimize the drift in error rates of recalibrations and coherent errors, and (ii) extend the duration of measurements. Based on this, all three CX gates for the 3-bit Boolean gates of CALA-*n* need to be replaced by three ECR gates.

It is noteworthy that: (i) if IBM quantum system decided to replace the CX or ECR gate with other double-qubit gates, e.g., the controlled-Z (CZ) gate, or (ii) if CALA-*n* utilized on other quantum systems, e.g., Google and Rigetti, then the three CX gates for all 3-bit Boolean gates need to match the double-qubit native gates of that quantum systems. Table 7 shows how the IBM quantum system transpiles the single-qubit Clifford+T gates to its supported single-qubit native gates {I, X, $\sqrt{X}$, RZ}, and Figure 11 illustrates a transpilation from the double-qubit Clifford gates to the IBM-supported double-qubit native gate {ECR} with the `ibm_brisbane` QPU.

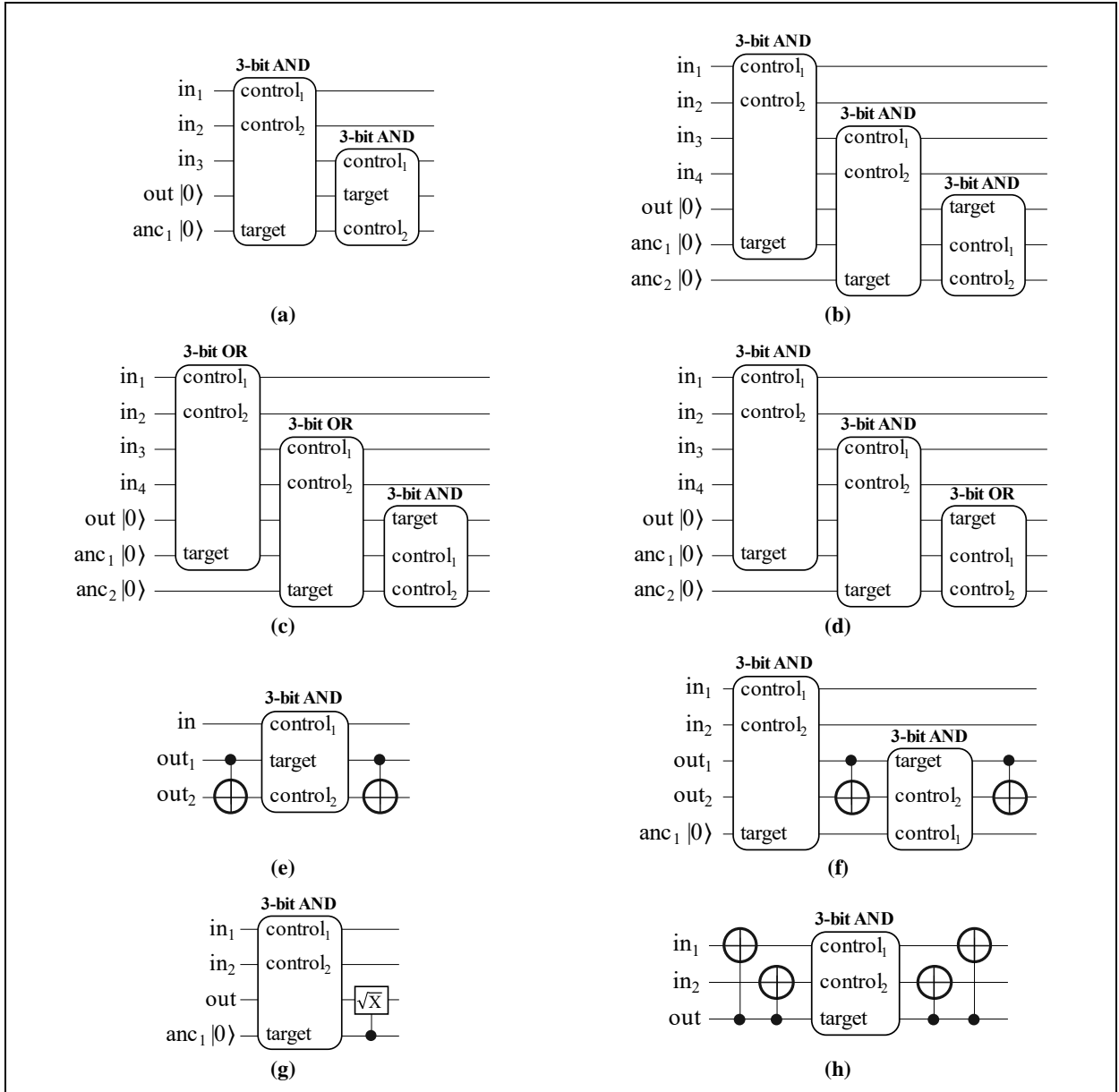

**Figure 10.** Schematics of $n$-bit gates and structures of CALA-$n$, where $3 \leq n \leq 5$ qubits: **(a)** 4-bit AND gate with $m = 1$, **(b)** 5-bit AND gate with $m = 2$, **(c)** 5-bit OR-AND-OR gate (as a POS structure) with $m = 2$, **(d)** 5-bit AND-OR-AND gate (as a SOP structure) with $m = 2$, **(e)** 3-bit Fredkin gate with $m = 0$, **(f)** 4-bit Fredkin gate with $m = 1$, **(g)** 3-bit controlled-$\sqrt{X}$ gate with $m = 1$, which is constructed from the 2-bit controlled-$\sqrt{X}$ gate shown in Figure 4(a), and **(h)** 3-bit Miller gate with $m = 0$. Note that (i) other 3-bit and 4-bit Boolean gates can be similarly constructed as in (a) and (b), (ii) different $n$-bit structures can be similarly constructed as in (c) and (d), (iii) 3-bit controlled-$\sqrt{X}^{\dagger}$ gate can be similarly constructed as in (g) by replacing $\sqrt{X}$ with $\sqrt{X}^{\dagger}$, which is constructed from the 2-bit controlled-$\sqrt{X}^{\dagger}$ gate shown in Figure 4(b), (iv) all ancilla (anc) qubits are initially set to the $|0\rangle$ state, and (v) $m$ is the total number of utilized anc for a gate.

**Table 7.** Transpilation of single-qubit Clifford+T gates to the single-qubit native gates {I, X, $\sqrt{X}$, RZ} supported by IBM QPUs of 127 qubits.

| Single-qubit Clifford+T gates | ➔ IBM single-qubit native gates |
|:---:|:---:|
| I | I |
| X | X |
| Y | {RZ($\pi$) X} |
| Z | RZ($\pi$) |
| H | {RZ($\pi$/2) $\sqrt{X}$ RZ($\pi$/2)} |
| $\sqrt{X}$ | $\sqrt{X}$ |
| $\sqrt{X}^\dagger$ | {$\sqrt{X}$ X} |
| S | RZ($\pi$/2) |
| S$^\dagger$ | RZ($-\pi$/2) |
| T | RZ($\pi$/4) |
| T$^\dagger$ | RZ($-\pi$/4) |

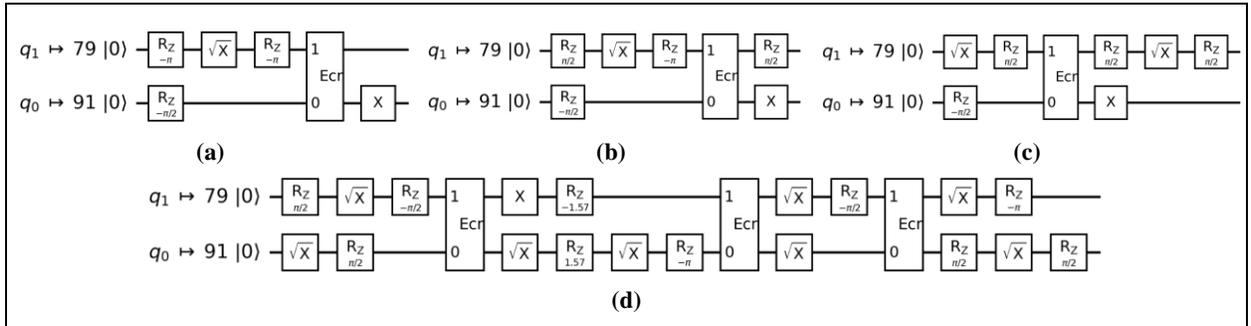

**Figure 11.** Transpilation of double-qubit Clifford gates to the native gates {I, X, $\sqrt{X}$, RZ, ECR} supported by IBM QPUs of 127 qubits: **(a)** the transpiled CX Clifford gate, **(b)** the transpiled CY Clifford gate, **(c)** the transpiled CZ Clifford gate, and **(d)** the transpiled SWAP Clifford gate. After transpilation with `ibm_brisbane` QPU, the two qubits ($q_0$ and $q_1$) of a double-qubit Clifford gate are routed to two neighboring physical qubits of indices 91 and 79, respectively. Note that different IBM QPUs of 127 qubits may generate different numbers and arrangements of $\sqrt{X}$, X, and RZ gates.

The 127 physical qubits of all IBM QPUs are arranged onto heavy-hex layouts (as hexagonal dense-connected lattices), with a limited $n$ physical qubits connectivity ($2 \leq n \leq 4$), as two-dimensionally illustrated in Figure 12 [8, 9]. To design cost-effective $n$-bit gates of CALA-$n$ ($2 \leq n \leq 5$), an I-shape of neighboring physical qubits arrangement is chosen to fulfill our research demands, as indicated by the red block in Figure 12. This I-shape represents the union of three linear arrangements, and each linear arrangement consists of three physical qubits. Hence, the I-shape can hold up to three 3-bit gates of CALA-$n$, i.e., each linear arrangement holds one 3-bit gate.

As discussed in the previous subsection, because the 3-bit gates are the fundamental gates of CALA-*n*, then three 3-bit gates (in an I-shape) can straightforwardly construct various *n*-bit gates of CALA-*n*. Therefore, such a straightforward construction fulfills our research demands by not requiring any SWAP gate among the physical qubits of an IBM QPU, through mapping the target qubit in the middle between the two control qubits. However, additional *m* ancilla qubits may be required for a 5-bit gate of CALA-*n*, as shown in Figure 13(a)

On the one hand, CALA-*n* overcomes the limited layout connectivity of an IBM QPU for greater than four neighboring physical qubits. On the other hand, if the IBM quantum system provided quantum layouts in the formation of square lattices, as in Google QPUs [23, 47], then a 9-bit gate of CALA-*n* can be straightforwardly constructed, as demonstrated in Figure 13(b). Subsequently, the previously constructed *n*-bit gates and structures of CALA-*n* (shown in Figure 10) can cost-effectively be mapped to the I-shape layout of an IBM QPU, as illustrated in Figure 14.

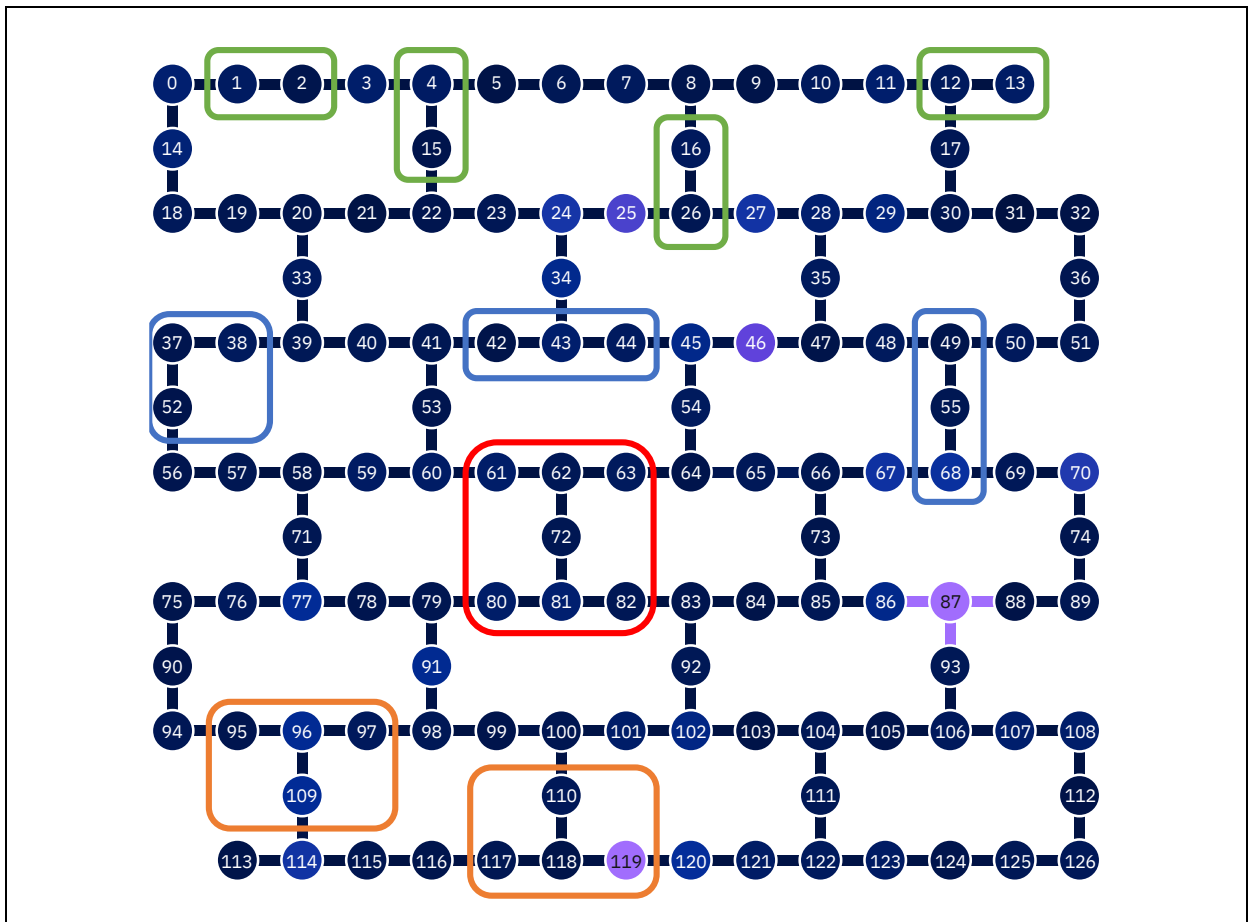

**Figure 12.** The layout map for all IBM QPUs of 127 qubits (including `ibm_brisbane`) [8, 9]: the green blocks are linear arrangements of any two neighboring physical qubits, the blue blocks are linear (or curved) arrangements of any three neighboring physical qubits, the orange blocks are joint linear-curved arrangements of any four neighboring physical qubits, and the red block is the I-shape of three linear arrangements (each has three qubits). The I-shape indicates the desired physical qubits arrangement that fulfills our research demands in designing cost-effective *n*-bit gates of CALA-*n*, where $3 \leq n \leq 5$ qubits.

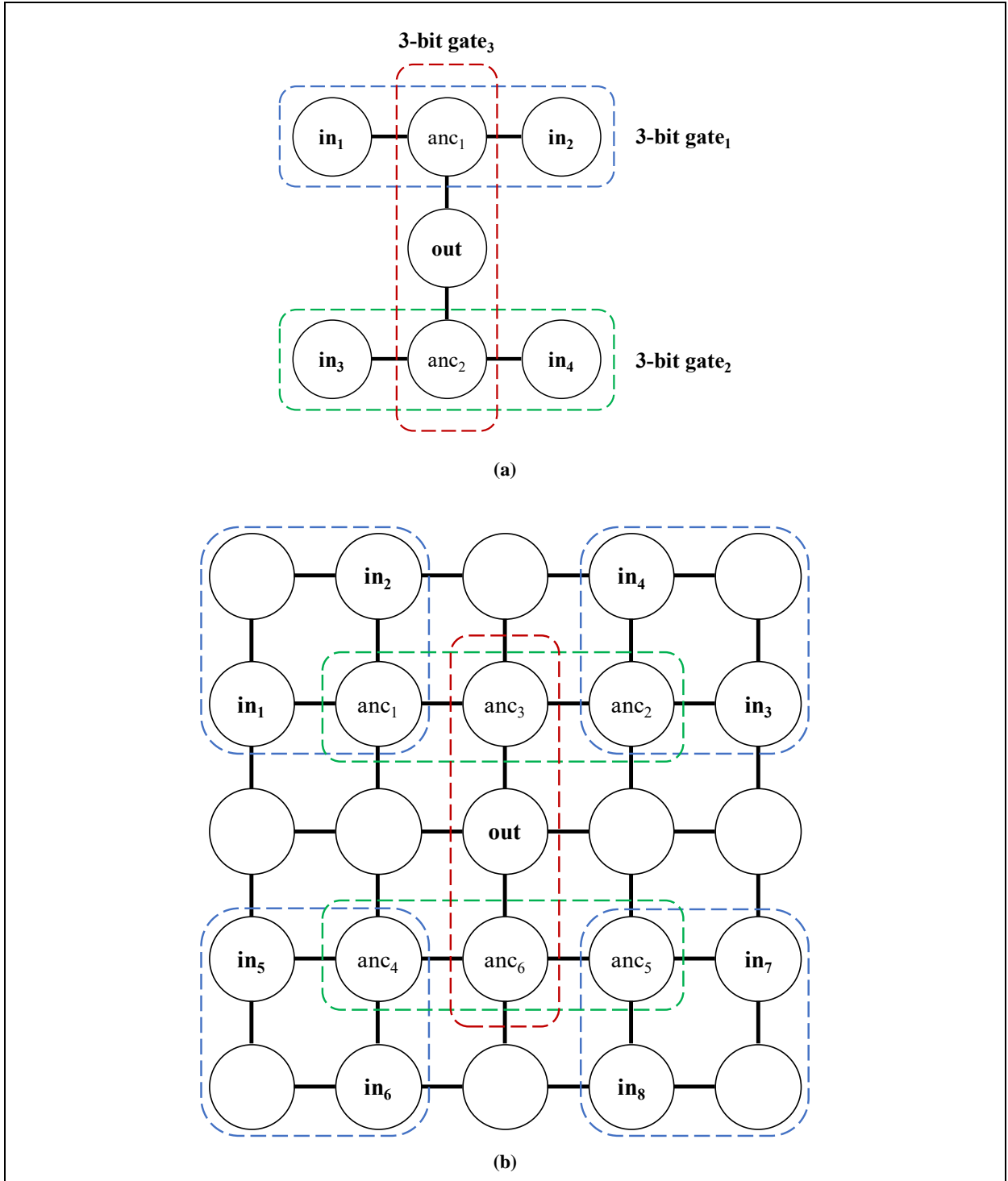

**Figure 13.** Schematics of implementing cost-effective *n*-bit gates of CALA-*n*, without adding SWAP gates to their final transpiled quantum circuits: **(a)** a 5-bit gate (four inputs, one output, and $m = 2$) utilizes three 3-bit gates using the I-shape layout of an IBM QPU, and **(b)** a 9-bit gate (eight inputs, one output, and $m = 6$) utilizes seven 3-bit gates using the square-lattice layout of a Google QPU [23, 47], where *m* is the total number of ancilla (anc) qubits. Note that CALA-*n* is a quantum library built from Clifford+T gates and mainly depends on the geometrical layout of a QPU.

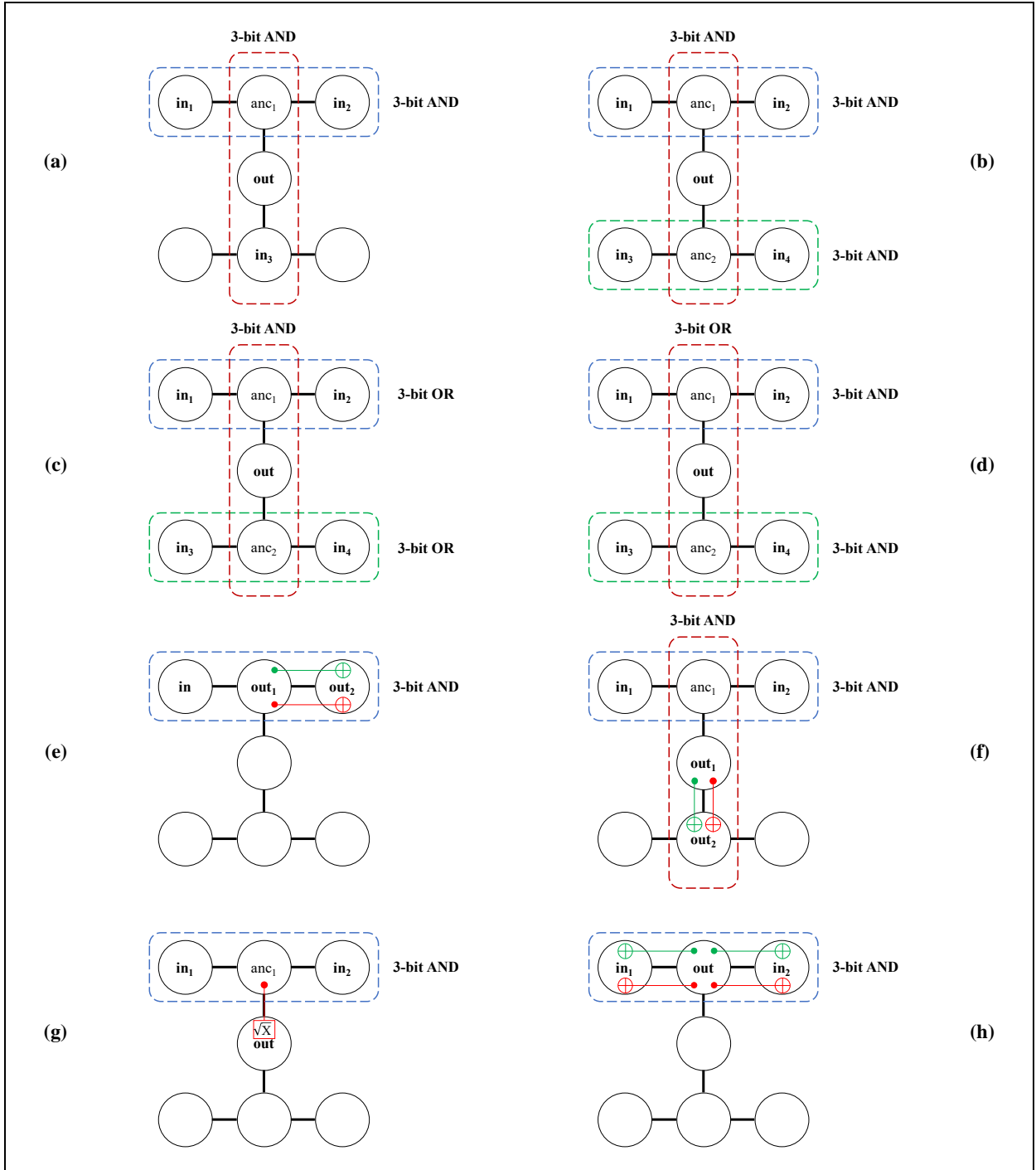

**Figure 14.** Schematics of *n*-bit gates and structures of CALA-*n* for cost-effectively mapping into the I-shape layout of an IBM QPU without utilizing any SWAP gate, where $3 \leq n \leq 5$ qubits: **(a)** 4-bit AND gate with $m = 1$, **(b)** 5-bit AND gate with $m = 2$, **(c)** 5-bit OR-AND-OR gate (as a POS structure) with $m = 2$, **(d)** 5-bit AND-OR-AND gate (as a SOP structure) with $m = 2$, **(e)** 3-bit Fredkin gate with $m = 0$, **(f)** 4-bit Fredkin gate with $m = 1$, **(g)** 3-bit controlled-$\sqrt{X}$ gate with $m = 1$, which is constructed from the 2-bit controlled-$\sqrt{X}$ gate shown in Figure 4(a), and **(h)** 3-bit Miller gate with $m = 0$. Note that (i) all CX gates in green are performed before their inbounded 3-bit AND gates, (ii) all controlled-$\sqrt{X}$ and CX gates in red are performed after their inbounded 3-bit AND gates, (iii) all ancilla (anc) qubits are initially set to the $|0\rangle$ state, and (iv) *m* is the total number of utilized anc.

## 3 RESULTS

As a proof of concept for building cost-effective *n*-bit gates of CALA-*n* for IBM QPUs, different *n*-bit gates are built and then transpiled with the `ibm_brisbane` QPU, where $3 \leq n \leq 5$ qubits. The quantum circuits of these *n*-bit gates are constructed using two approaches:

1. *The standard approach*: by using the standard quantum gates, e.g., *n*-bit Toffoli, Feynman (CX), Fredkin (controlled-SWAP), and controlled-$\sqrt{X}$ (controlled-V) gates. Most of these gates are non-Clifford+T gates as well as non-native gates in many IBM QPUs. For that, further decompositions are required for their final transpiled quantum circuits.
2. *The Bloch sphere approach*: by using the *n*-bit gates of CALA-*n*, as discussed in the previous section and shown in Figure 10. All *n*-bit gates of CALA-*n* are entirely constructed from Clifford+T gates, which are mostly native gates in many IBM QPUs. For that, no decomposition is required for their final transpiled quantum circuits.

After transpilation with `ibm_brisbane` QPU, the final transpiled quantum circuits of these *n*-bit gates (for the two approaches) are evaluated based on the total number of generated native gates. In this paper, the total number of generated native gates is considered as the quantum cost. Such that, a transpiled quantum circuit with a higher quantum cost is considered as a cost-expensive gate, while a transpiled quantum circuit with a lower quantum cost is considered as a cost-effective gate. Thereby, we concluded that all *n*-bit gates constructed from the standard approach are always cost-expensive gates, while all *n*-bit gates constructed from the Bloch sphere approach are always cost-effective gates, which are the *n*-bit gates of CALA-*n*.

The IBM Quantum Platform was used to design and evaluate the *n*-bit gates (for the two approaches) with `ibm_brisbane` QPU, for 1024 resampling times, which are the so-called "shots" [48]. However, the technical specifications of T1, T2, frequency, anharmonicity, median errors of native gates, speed of transpilation, number of pulses, and the usage load of this QPU are not considered for the purpose of our research and experiments.

For the standard approach, the transpilation automatically maps the qubits of an *n*-bit gate to the physical qubits of `ibm_brisbane` QPU. If these mapped physical qubits are not directly connected to each other, then a number of SWAP gates are added. In contrast, for the Bloch sphere approach, we manually indicate which qubits of an *n*-bit gate should be mapped to a list of physical qubits in the I-shape layout of this QPU; hence, SWAP gates are not required. In such a way, we ensure to have transpiled cost-effective quantum circuits for all *n*-bit gates of CALA-*n*. Note that all *n*-bit gates of CALA-*n* utilize the I-shape layout for the physical qubits of the indices {61, 62, 63, 72, 80, 81, 82}, as indicated by the red box shown in Figure 12. Table 8 states the resultant quantum costs for the *n*-bit gates after the transpilation, for both approaches (the standard and the Bloch sphere).

**Table 8.** The resultant quantum cost (QC) of different $n$-bit gates after transpiling to the native gates (X, $\sqrt{X}$, RZ, and ECR) of `ibm_brisbane` QPU, using the standard approach and our proposed Bloch sphere approach, where $3 \leq n \leq 5$ qubits.

| $n$-bit gates | The standard approach (using standard quantum gates) | | | | | The Bloch sphere approach (using Clifford+T gates) | | | | |
|---|---|---|---|---|---|---|---|---|---|---|
| | X | $\sqrt{X}$ | RZ | ECR | QC | X | $\sqrt{X}$ | RZ | ECR | QC |
| 3-bit Toffoli (AND) | 2 | 18 | 28 | 7 | **55** | 0 | 14 | 20 | 3 | **37** |
| 4-bit Toffoli (AND) | 3 | 39 | 60 | 16 | **118** | 3 | 19 | 29 | 6 | **57** |
| 5-bit Toffoli (AND) | 18 | 149 | 220 | 65 | **452** | 4 | 31 | 46 | 9 | **90** |
| 5-bit OR-AND-OR (POS) | 6 | 66 | 98 | 26 | **196** | 4 | 31 | 46 | 9 | **90** |
| 5-bit AND-OR-AND (SOP) | 13 | 59 | 97 | 27 | **196** | 4 | 31 | 46 | 9 | **90** |
| 3-bit Fredkin (controlled-SWAP) | 3 | 26 | 38 | 10 | **77** | 0 | 12 | 23 | 5 | **40** |
| 4-bit Fredkin (controlled-SWAP) | 7 | 56 | 80 | 20 | **163** | 0 | 28 | 44 | 8 | **80** |
| 3-bit controlled-$\sqrt{X}$ (controlled-V) | 5 | 30 | 45 | 11 | **91** | 0 | 19 | 31 | 4 | **54** |
| 3-bit Miller (distance) | 3 | 36 | 50 | 13 | **102** | 2 | 16 | 25 | 7 | **50** |

From Table 8, the following key points were observed between the standard approach and the Bloch sphere approach after transpiling all $n$-bit gates:

1. The number of ECR gates from the standard approach is (at minimum) twice the number of ECR gates from the Bloch sphere approach, due to the reason that SWAP gates are never utilized in the Bloch sphere approach, i.e., in all $n$-bit gates of CALA-$n$.

2. From the Bloch sphere approach, all 5-bit gates of CALA-$n$ have the same quantum costs, due to the reason that they share the same quantum layout mapping (see Figure 14) as well as the same symmetrical quantum circuit (see Figure 6) of permutative four Clifford+T ($\theta$) gates and customized AX$_2$ gates (see Table 6). For instance, the quantum cost of the 5-bit AND gate is equal to that of the 5-bit OR gate, since the OR's $\theta_4$ gate (as T) is merged with the AX$_2$ gate (as Z) to form one RZ($5\pi/4$) gate after the transpilation.

3. The quantum costs from the standard approach are always higher than that of the Bloch sphere approach. For instance, approximately, (i) the quantum cost of the 3-bit Miller gate from the standard approach is twice as high as that of the Bloch sphere approach, and (ii) the quantum cost of the 5-bit Toffoli gate from the standard approach is five times higher than that of the Bloch sphere approach!

Accordingly, our proposed Bloch sphere approach as the geometrical design tool successfully generates various cost-effective *n*-bit gates built from Clifford+T gates, collectively as CALA-*n* quantum library. For the IBM quantum system, CALA-*n* can provide a broad range of various cost-effective quantum gates and Boolean operators for practical applications in the fields of Boolean oracular algorithms, digital logic circuits, robotics, machine learning, just to name a few. Our future work will concentrate on cost-effectively re-building CALA-*n* based on various native gates and layouts of other superconducting quantum systems, such as Google and Rigetti.

## 4 CONCLUSION

The standard 3-bit Toffoli gate is mainly used to construct different standard gates, such as Peres, inverse Peres, Fredkin, and *n*-bit Toffoli gates, where $n > 3$ qubits. These standard gates are entirely constructed from the mathematical calculations of unitary matrices. However, these standard gates are non-native gates in many IBM quantum computers. Native gates are operationally supported by a quantum computer. For that, a standard gate is decomposed into a set of native gates, as well as its qubits are routed to be mapped into the physical qubits in the architecture (layout) of a quantum computer. However, the routing process inserts a number of SWAP gates to connect the non-neighboring physical qubits together. In IBM terminologies, the decomposition, routing, and mapping processes are termed "transpilation". After transpilation, we noticed that different standard gates have higher quantum costs, i.e., they are cost-expensive gates. The quantum cost, in this paper, denotes the total number of generated native gates for the transpiled quantum circuit of a standard gate. In our research, we mainly focus on constructing various sets of cost-effective gates, which have similar quantum operations to their equivalent cost-expensive standard gates. The cost-effectiveness of our gates comes from the utilization of Clifford+T gates (before transpilation) as well as the non-utilization of SWAP gates (after transpilation), since a set of Clifford+T gates is mainly utilized by many IBM quantum computers as native gates, e.g., T, T$^\dagger$, and Feynman (CX) gates.

In this paper, we introduce a new methodology for constructing cost-effective gates using the Bloch sphere approach as the geometrical design tool, instead of using the mathematical calculations of unitary matrices. Based on this Bloch sphere approach and the Clifford+T gates, these cost-effective gates have one generalized architecture, which mainly depends on the limited layout connectivity of IBM quantum computers, for up to five physical qubits. Various cost-effective *n*-bit gates are constructed to perform Boolean operations, switching functionalities, and square rooting. We termed these cost-effective *n*-bit gates as the "Clifford+T-based Architecture of Layout-Aware *n*-bit gates (CALA-*n*)", where $2 \leq n \leq 5$ qubits. In our work, the `ibm_brisbane` quantum computer of limited layout connectivity is chosen as a proof of concept for CALA-*n*. Different standard *n*-bit gates and their equivalent gates of CALA-*n* were examined and evaluated with this quantum computer, and it was concluded that all gates of CALA-*n* always have lower quantum costs than those of standard *n*-bit gates after transpilation. In conclusion, CALA-*n* as a quantum library of cost-effective *n*-bit gates can be employed to build Boolean oracles, implement quantum oracular algorithms, and enhance efficient transpilation based on the layouts of IBM quantum computers. Moreover, CALA-*n* can provide a broad range of practical applications in the fields of digital logic synthesizers, computer vision, machine learning, just to name a few, in the quantum domain.